\documentclass[9pt,twocolumn,twoside]{pnas-new}

\usepackage{subfig}

\usepackage{xcolor}
\definecolor{highlight}{rgb}{0,0,0}

\templatetype{pnasresearcharticle} 

\title{Universal Fermi-surface anisotropy renormalization for interacting Dirac fermions with long-range interactions}

\author[a,b]{Jia Ning Leaw}
\author[a,b]{Ho-Kin Tang}
\author[b]{Maxim Trushin}
\author[c]{Fakher F. Assaad}
\author[a,b,d,1]{Shaffique Adam}

\affil[a]{Department of Physics, National
University of Singapore, 2 Science Drive 3, 117551, Singapore}
\affil[b]{Centre for Advanced 2D Materials, National University of Singapore, 6 Science Drive 2, Singapore 117546}
\affil[c]{{Institut f\"ur Theoretische Physik und Astrophysik, Universit\"at W\"urzburg, Am Hubland, D-97074 W\"urzburg, Germany.}}
\affil[e]{Yale-NUS College, 16 College Avenue West, 138614, Singapore}

\leadauthor{Leaw} 

\significancestatement{Electrons in a two-dimensional metal have a one-dimensional contour in momentum space, called the Fermi surface, that separates the occupied and unoccupied energy levels.  For realistic materials the Fermi surface is usually anisotropic, while theoretical models often assume rotationally symmetric Fermi surfaces.  In this work, we address a simple but fundamental question: Given a specified anisotropic Fermi surface for a non-interacting two-dimensional electron gas, what happens to this anisotropy in the presence of electron-electron interactions? Remarkably, only for Dirac electrons with long-range Coulomb interactions there is a universal square-root renormalization of the anisotropy, independent of the interaction strength.  This prediction can be tested in a variety of experimental systems including graphene, topological insulators, organic-layered compounds, twisted bilayer graphene and others. }

\authorcontributions{SA conceived this project.  HKT and FFA developed and optimized the quantum Monte Carlo codes.  JNL and MT developed the analytical results under the guidance of SA.  All authors discussed the results and wrote the paper.}
\authordeclaration{The authors declare that they have no competing financial interests.}
\correspondingauthor{\textsuperscript{1}To whom correspondence should be addressed. E-mail: shaffique.adam@yale-nus.edu.sg}


\begin{abstract}
Recent experimental~\cite{Jo2017} and numerical~\cite{Ippoliti2017} evidence suggest an intriguing universal relationship between the Fermi surface anisotropy of the non-interacting parent two-dimensional electron gas and the strongly correlated composite Fermi liquid formed in a strong magnetic field close to half-filling.  Inspired by these observations, we explore more generally the question of anisotropy renormalization in interacting 2D Fermi systems.  Using a recently developed~\cite{Tang2018} non-perturbative and numerically-exact projective quantum Monte Carlo simulation as well as other numerical and analytic techniques, only for Dirac fermions with long-range Coulomb interactions do we find a universal square-root decrease of the Fermi-surface anisotropy.  For the $\nu=1/2$ composite Fermi liquid, this result is surprising since a Dirac fermion ground state~\cite{Son2015} was only recently proposed as an alternative to the usual HLR state~\cite{Halperin1993}.  The importance of the long-range interaction, expected for Dirac systems~\cite{Adam2007}, is also consistent with recent transport measurements~\cite{Pan2017}.  Our proposed universality can be tested in several anisotropic Dirac materials including graphene, topological insulators~\cite{Kim2012}, organic conductors~\cite{Hirata2016}, and magic-angle twisted bilayer graphene~\cite{Bistritzer2011}.
\end{abstract}

\dates{This manuscript was compiled on \today}
\doi{\url{www.pnas.org/cgi/doi/10.1073/pnas.XXXXXXXXXX}}

\begin{document}

\maketitle
\thispagestyle{firststyle}
\ifthenelse{\boolean{shortarticle}}{\ifthenelse{\boolean{singlecolumn}}{\abscontentformatted}{\abscontent}}{}

\dropcap{A} plethora of quantum Hall ground states are observed when the two-dimensional electron gas is placed in a strong magnetic field.  Unique among these states is when the system is tuned to half-filling; here rather than observing an insulator as is typical, instead a metallic state is observed.  This compressible phase called a ``composite Fermi liquid” is believed to emerge because at the mean field level, the magnetic flux attached to each composite fermion exactly cancels the external magnetic field~\cite{sarma2008,Prange1989}.  The Fermi surface properties of this metallic state have long been explored.  A very recent experiment at Princeton measured how the anisotropy of this strongly correlated Fermi surface along the high-symmetry axis (parameterized by $\eta$, defined below) was related to that of the original Fermi surface $\eta_0$ of the non-interacting electrons in the absence of an external magnetic field~\cite{Jo2017}.  Before performing the experiment, they surveyed leading theorists on what they expected to observe, obtaining a range of answers:  since the composite Fermi liquid was a universal strongly correlated ground state, in which the kinetic energy of the non-interacting bands were quenched, some expected no Fermi surface anisotropy in the composite Fermi liquid; others expected the state to have the same anisotropy as the non-interacting bands from which the state emerged.  Remarkably, the experiment observed that $\eta = \sqrt{\eta_0}$ i.e. the interacting state was always more isotropic and universally so. \\

\noindent This experiment motivated us to ask a more fundamental question.  What is the relationship between the Fermi surface anisotropy of an interacting Fermi liquid given a fixed anisotropy of the non-interacting bands i.e. we are interested not only in the effective composite Fermi liquid at $\nu=1/2$, but the interacting 2D Fermi liquid in general.  To illustrate that the answer is not obvious consider the following.  One might expect that interactions enhance the anisotropy.  Anisotropy can be thought of as broken rotational symmetry, and it was predicted long ago that the exchange interaction can enhance the splitting between broken symmetry states~\cite{Ando1974}.   In the quantum Hall context, interaction-enhanced Zeeman splitting has been seen experimentally for the graphene integer quantum Hall effect~\cite{Song2010} and more recently, an interaction-induced spontaneous symmetry breaking of nematic phases~\cite{Feldman2016}.   \\

\noindent In other contexts, interactions wash away non-universal particularities of the non-interacting model flowing to a universal interacting fixed point (the universal $\nu = 1/3$ Laughlin state observed in different parent materials and with different confining potentials leading to different effective interactions, is but one example).  Experimentally, the observed sequence of fractional quantum Hall plateaus in graphene suggest that the strongly interacting ground state partially restores the spin and valley splitting of the non-interacting system~\cite{Young2012}.  For the specific case of $\nu = 1/2$, using a Gaussian approximation for the electron-electron interactions, Ref.~\cite{Yang2013} found analytically that interactions always make the composite Fermi liquid more isotropic, but in a non-universal way where $\eta/\eta_0$ could take on values between 0 and 1 depending on the length scale of the Gaussian.  Other calculations in specific models~\cite{Balagurov2000,Balram2016} suggest no change in the Fermi surface anisotropy; while yet others show a non-universal decrease in anisotropy~\cite{Yang2012,Murthy2013}. \\

\noindent We are not the first to ask the question about the many-body renormalization of anisotropic Fermi surfaces.  Back in 1960, Kohn and Luttinger~\cite{Kohn1960} argued that the standard diagrammatic perturbation theory to account for electron-electron interactions failed when considering anisotropic Fermi surfaces.  Moreover, the effect of correlations was understood to be non-universal where the anisotropy renormalization depends on material specific parameters like the effective mass, carrier density and dielectric substrate.  Below we reproduce the leading order term for ``Schr\"odinger electrons" where the bare band dispersion is of the usual parabolic energy form to illustrate how this non-universality arises for generic band structures.  However, and remarkably, we find that for Dirac fermions (e.g. graphene) with bare Fermi velocity anisotropy, in the presence of a long-range Coulomb interaction, there is a universal relationship $\eta = \sqrt{\eta_0}$ that does not depend on any of the material-specific parameters mentioned above.  We emphasize that the long-range nature of the Coulomb potential is essential: our analytical and numerical results also suggest that Dirac fermions with only contact interactions retain the anisotropy of the original non-interacting system.  We find that the chirality of the Dirac bands and the long-range interaction are both necessary to obtain the square-root anisotropy.\\

\begin{figure}
\center
\includegraphics[width=0.5\textwidth]{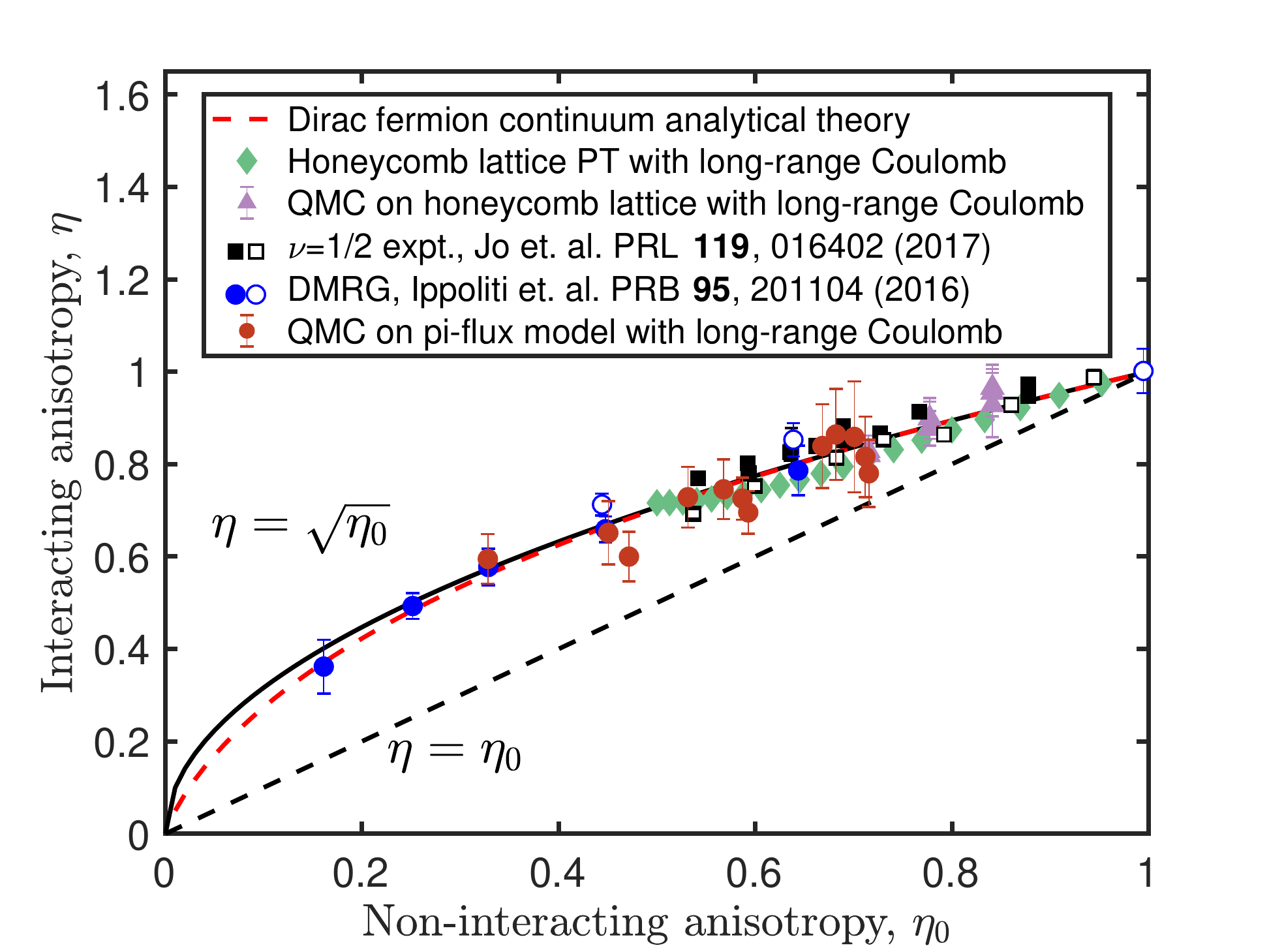}
\caption{Universal decrease of Fermi-velocity anisotropy for long-range interacting Dirac fermions.  Analytical perturbation theory and Hartree-Fock results for Dirac fermions (red dashed line) show a square-root decrease of the Fermi velocity anisotropy.  Lattice perturbation theory on the honeycomb lattice (green diamonds){\color{highlight},} the non-perturbative projective Quantum Monte Carlo {\color{highlight} on honeycomb lattice} (purple triangles) {\color{highlight} and $\pi$-flux model (red circles)} are {\color{highlight} all} consistent with this square-root behaviour.  Also shown are recent experiments (black squares) and density matrix renormalization group calculations (blue circles) for the $\nu = 1/2$ quantum Hall state. We find that both the chirality of the Dirac bands and a long-range interaction potential are necessary for this universal decrease in anisotropy. \label{fig:Fig1}}
\end{figure}

\begin{figure*}
\center
\subfloat[\label{fig:Fig4a}]{\includegraphics[width=0.5\linewidth]{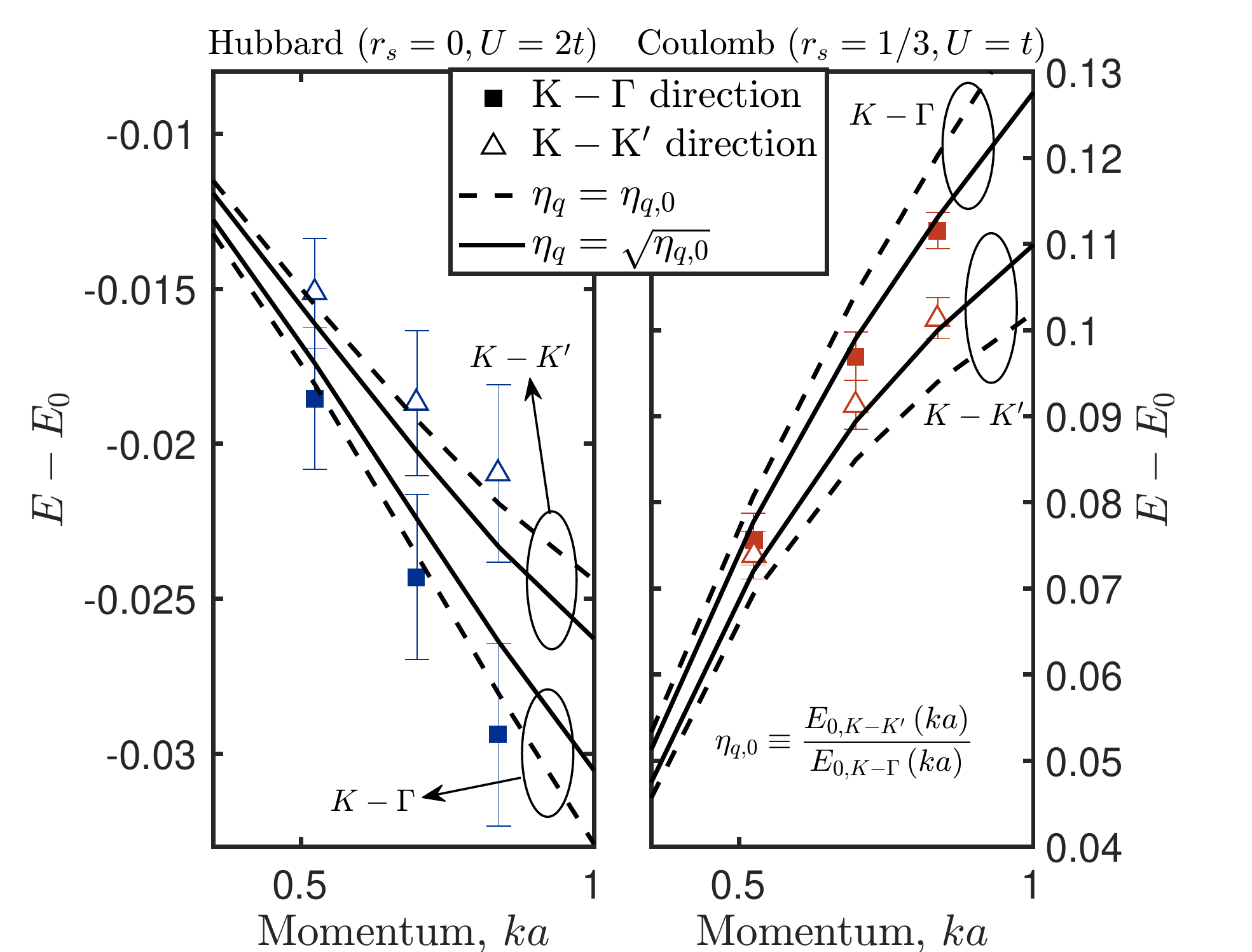}}
\subfloat[\label{fig:Fig4b}]{\includegraphics[width=0.5\linewidth]{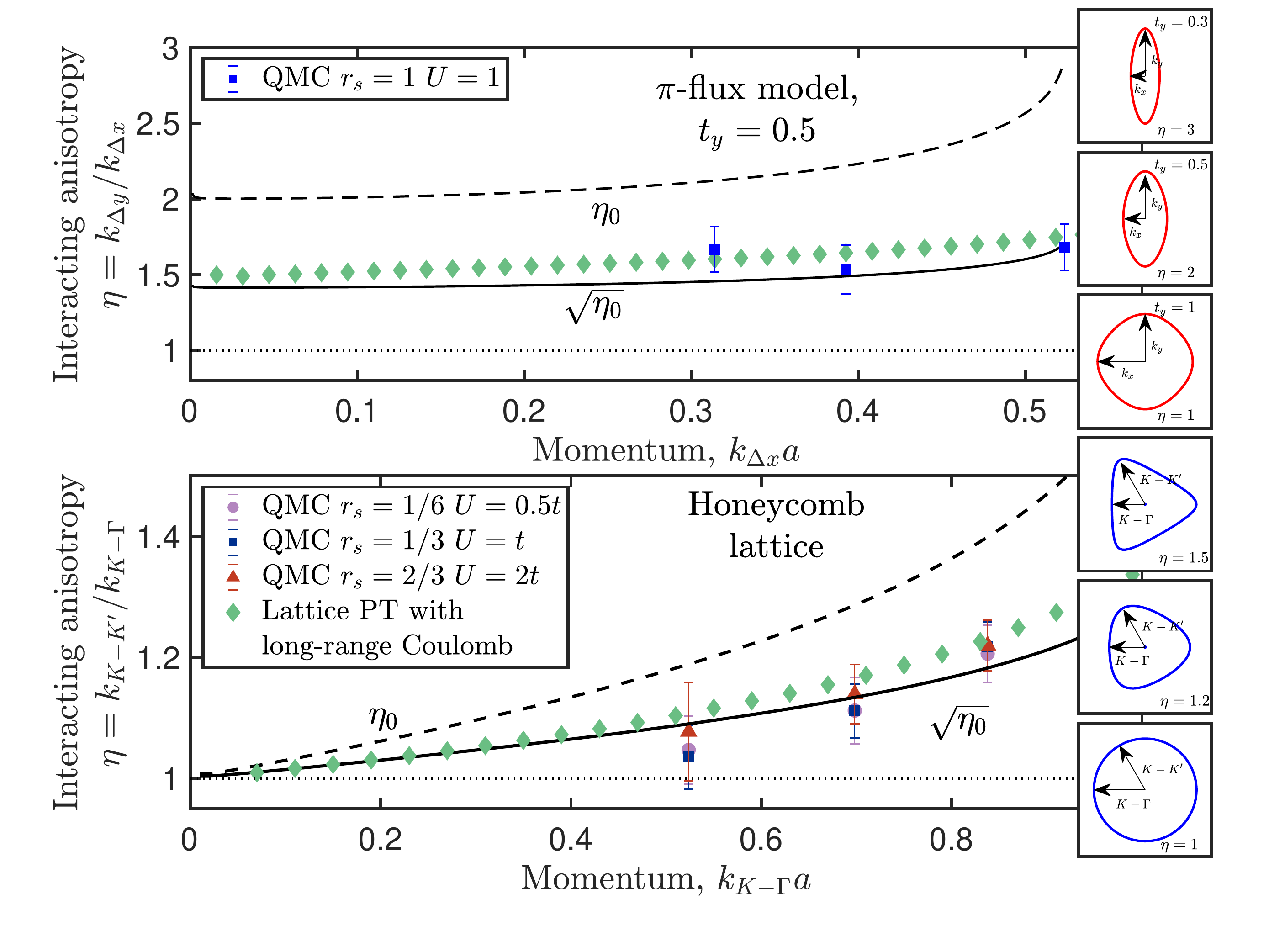}}
\caption{Non-perturbative quantum Monte Carlo simulations on a {\color{highlight} $\pi$-flux model  and the honeycomb lattice both} show universal square-root anisotropy renormalization with long-range Coulomb interactions, but not for Hubbard interactions. (a) QMC data {\color{highlight} on honeycomb lattice} with onsite Hubbard interaction (left panel) and long-range Coulomb interaction (right panel). The energy renormalization $E-E_0$ of the quasiparticle is obtained for $K-\Gamma$ direction (squares) and $K-K'$ direction (triangles).  Solid lines indicate what one would expect for a square-root renormalization, while dashed lines show no anisotropy renormalization.  (b) The interacting theory anisotropy as a function of momentum away from the Dirac point.  The {\color{highlight} $\pi$-flux model has an anisotropy that is almost constant for small momentum down to the Dirac point, while the} honeycomb lattice is isotropic as momentum vanishes but becomes more anisotropic at larger momentum.  Both the non-perturbative QMC and the lattice perturbation theory show a clear square-root anisotropy renormalization for long-range Coulomb interactions. \label{fig:Fig4}}
\end{figure*}

\noindent In the presence of Coulomb interactions, the properties of correlated Dirac fermions are governed by two very different fixed points~\cite{Tang2018}.  There is a stable ``weak-coupling” fixed point determined entirely by the long-range Coulomb tail, which flows at low energies to a non-interacting Lorenz invariant theory (where the electron Fermi velocity is equal to the speed of the light).  Then there is also an unstable ``strong-coupling” Gross-Neveu fixed point controlled by the contact part of the Coulomb interaction and associated with the transition to a Mott insulating antiferromagnet at half-filling.  While ultimately unstable, the proximity to the strong coupling fixed point is largely responsible for most experimentally observable properties of Dirac fermions.  Paradoxically, the closer to the ``strong-coupling” fixed point the flow of Dirac fermion under interactions starts, the more it behaves like the non-interacting theory (with typical Fermi velocities about two or three orders of magnitude slower than the speed of light)~\cite{Tang2018}.   The dichotomy between the long-range and contact part of the Coulomb potential for Dirac fermions was also discussed in Ref.~\cite{Banerjee2018}.   Since this particular velocity renormalization influences all points on the Fermi surface equally, it drops out when calculating the anisotropy renormalization that we are concerned with here. \\ 

\noindent To build some intuition into our results, we begin by considering the Hartree-Fock approximation and first-order perturbation theory (for the case we consider, both these considerations give identical results, see Methods section below).  This is the leading-order many-body result as expanded directly in the bare interaction. We first observe that for Dirac fermions, the contact interaction does not renormalize the Fermi velocity (see e.g. Ref.~\cite{Giuliani_CMP:2009} and Methods section).  For the long-range Coulomb interaction, the dominant contribution comes from the filled hole bands.  Although the Fermi surface is vanishing at the Dirac point, the Fermi surface anisotropy is still well-defined. By first setting a finite Fermi level (defined by a Fermi momentum $k_F$) and then taking the limit $k_F\rightarrow 0$, the Fermi surface anisotropy converges to the Fermi velocity anisotropy $\eta_0 = v_x/v_y$, where the subscript denotes that this is the non-interacting system.  When the 2D Coulomb interaction is turned on, the effective single particle energy is corrected by the exchange correlation, which within the Hartree-Fock approximation 
is given by~\cite{Trushin2011}
\begin{equation}
E^{HF}(\mathbf{k})=-\int_{\Omega}\frac{d^2\mathbf{k}'}{(2\pi)^2}\frac{2\pi e^2}{\epsilon|\mathbf{k}-\mathbf{k}'|}\sin^2\left(\frac{\varphi_k-\varphi_k'}{2}\right),
\label{eq:DiracExchangeEnergy}
\end{equation}
where $e$ is the electron charge, $\epsilon$ is the dielectric relative permittivity,
and $\tan\varphi_k =  (v_y k_y)/(v_x k_x)$. The $\varphi_{k}$-dependent factor accounts for the chirality of
the Dirac fermions. \textcolor{highlight}{The Fermi velocity renormalization along the $x$-
and $y$- axes were calculated in Ref.~\cite{Dugaev2012}. Taking the
ratio of the velocity along the $x$-axis to the $y$-axis, we find the anisotropy
\begin{equation}
\eta =\frac{\mathrm{E}(1-\eta_0^2)-\mathrm{K}(1-\eta_0^2)}{\mathrm{K}(1-\eta_0^{-2})-\mathrm{E}(1-\eta_0^{-2})} \approx \sqrt{\eta_0}. \label{eq:DiracAnisotropy}
\end{equation}
Here $\mathrm{K}\left(z\right)$ and $\mathrm{E}\left(z\right)$ are the
complete elliptic integrals of the first and second kind.
By plotting this analytic result as the red curve in Fig. 1, we make
a crucial observation: the renormalized anisotropy is remarkably close
to the square-root of $\eta_{0}$.}  Since within Hartree-Fock and first-order perturbation theory the exchange energy contribution dominates over the non-interacting contribution, it is not surprising that the result is universal, nonetheless this behavior is unique (see Methods) to Dirac fermions in two-dimensions with a long-range Coulomb interaction.  We emphasize that chirality is essential for our result.  The chiral eigenfunctions of the Dirac fermions are invariant to the Hartree-Fock interaction and neglecting these, or even changing the winding number from 1 breaks the universality.  By contrast, particle-hole symmetry breaking terms do not couple to the chirality and preserve the universality (see Methods for details).  \\

\noindent While the perturbative approach is useful to understand qualitatively how the universality arises in the Dirac system, here we use a non-perturbative, numerically exact projective quantum Monte Carlo simulation appropriate for the strongly correlated nature of the ground state.  We use a honeycomb lattice with nearest neighbor hopping and at half-filling there is no fermion sign problem.  We are able to separately tune the short-range or Hubbard $U$ and long-range tail $r_s/r$ components of the Coulomb interaction (see Methods).  The non-interacting system contains Dirac cones at the high-symmetry $K$, $K'$ points in the Brillouin zone that are isotropic in the low-energy limit, but have trigonal warping away from the Dirac point naturally providing an anisotropic Fermi surface, whose renormalization with interactions we can explore.  Appropriate for the three-fold symmetry, we define $\eta_q$ as the ratio of the energy deviation $(E-E_0)_{K-K'}/(E-E_0)_{K-\Gamma}$ at a fixed momentum, where $K-K'$ and $K-\Gamma$ are the principal directions. \\

\noindent {\color{highlight} In addition to honeycomb lattice, we also perform projective quantum Monte Carlo on a $\pi$-flux model.  We consider on a square lattice the nearest-neighbor hopping $t_{ij}$ with a spatially varying phase, such that the product of phases of hopping integrals around a plaquette is $e^{i\pi} = -1$. Similar to honeycomb lattice, the $\pi$-flux model also has two Dirac points, which are situated at $k=(\pi/2,\pm \pi/2)$ in the Brillouin zone. To get anisotropic Dirac cones, we set the magnitude of the hopping integral along the $x$-axis to 1, while varying $t_y$, the magnitude of the hopping integral along the $y$-axis.  In contrast to the honeycomb lattice, the non-interacting dispersion remains anisotropic even as we approach the Dirac point.}

\noindent Our results are obtained from a finite-size scaling of the numerical data on lattice sizes up  to {\color{highlight}$32 \times 32$} unit cells.  The left panel of Fig.~\ref{fig:Fig4a} shows a typical case where short-range interactions are dominant ($r_s = 0; U = 2t$).  In this case the correlation induced velocity renormalization is very small, and the data are consistent with no renormalization of the anisotropy, $\eta_q = \eta_0$ within the error bars.  The right panel is typical for when the long-range tail dominates ($r_s = 1/3, U = t$).  The data shows that strongly correlated Dirac fermions with long-range Coulomb interactions have the universal $\eta_q = \sqrt{\eta_{q,0}}$.  {\color{highlight} Similar results are also shown for $\pi$-flux at $t_y=0.5$.}  To complement the quantum Monte Carlo, we calculate first order perturbation theory on the honeycomb lattice, allowing us to go to much bigger system sizes ($1500 \times 1500$ unit cells), showing again good agreement with the square-root renormalization of the anisotropy (see Methods for details).  The various numerical and analytical approaches all confirm that the long-range Coulomb potential gives a universal square-root supression of the bare anisotropy for interacting Dirac fermions, reminiscent of the experimental finding at the half-filled Landau level in Ref.~\cite{Jo2017}. \\  

\begin{figure}
\center
\includegraphics[width=0.95\columnwidth]{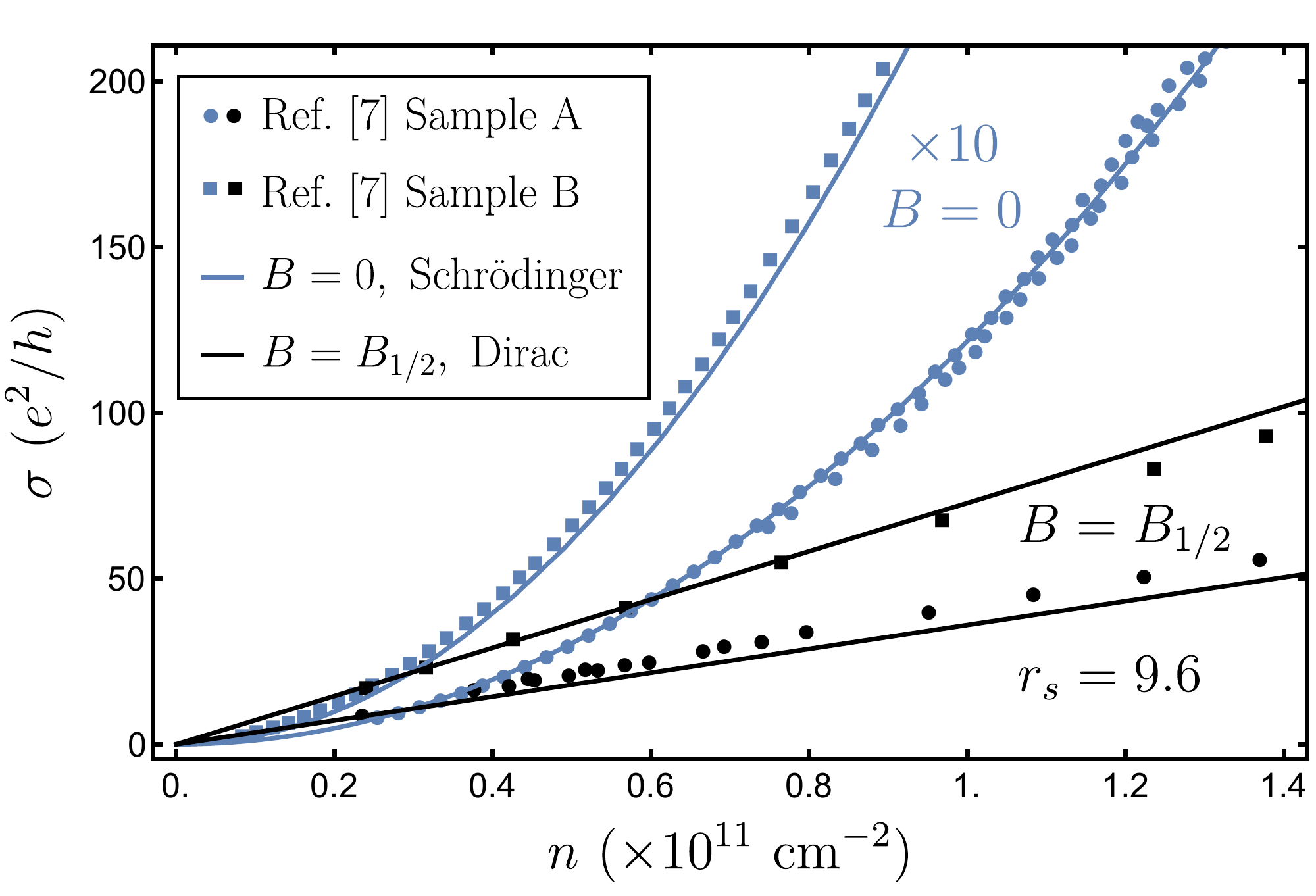}
\caption{The bare electron and composite fermion conductivities, $\sigma$, {\it vs.} fermion concentration, $n$, obtained for a given sample at zero magnetic field ($B=0$) and near half-filling ($B= B_{1/2}$). The circles and squares are the experimental data by Pan et al.~\cite{Pan2017} for two different samples.  The zero-field data shows a quadratic trend fitted by the Boltzmann conductivity for Schr\"odinger fermions with impurity concentrations $n_i = 1.9\times10^7\text{ cm}^{-2}$ and $9.6\times10^6\text{cm}^{-2}$ for samples A and B respectively.  The conductivity at half-filling for both samples is then obtained by the Boltzmann conductivity for Dirac fermions assuming a single Fermi velocity $v =\sqrt{v_xv_y}= 1.8\times 10^6$ cm/s.
\label{fig:FigVelocity}}
\end{figure}

\noindent  A natural interesting question is whether this is a mere coincidence or is an indicator for the predicted emergent Dirac fermion nature of the half-filled Landau level.  While answering this question definitively is well beyond the scope of the current work, which is entirely on the many-body renormalization of bare anisotropy in zero-field interacting 2D Dirac and Schr\"odinger systems, we mention one more tantalizing experimental finding in this context by Pan et al.~\cite{Pan2017} who established that the half-filled Landau level conductivity in high-mobility modulation-doped 2D structures manifests a linear-in-carrier density dependence (similar to that observed in graphene~\cite{Tan2007})  in contrast to the expected quadratic density dependence observed and predicted in the corresponding zero-field 2D conductivity of the usual Schr\"odinger system~\cite{DasSarma2015}.  If our speculations about a possible connection between our result and the emergent Dirac nature of the composite fermion liquid at half-filling are correct, then it also follows that the long-range nature of Coulomb interaction is an essential part of the physics here since we only find the square-root suppression of the bare anisotropy for long-range Coulomb interaction and not for the short-range contact interaction.\\

\noindent Shown in Fig.~\ref{fig:FigVelocity} is the conductivity for two of the samples reported in Ref.~\cite{Pan2017} both at $B=0$ (blue data) and close to half-filling (black data).  For unscreened long-range interactions, the conductivity $\sigma \sim (e^2/h) (n/n_i) (1/r_s^2)$, where $r_s =e^2/(\epsilon\hbar \sqrt{v_xv_y})$ is density independent for Dirac fermions but scales as $n^{-1/2}$ for Schr\"odinger fermions.  Assuming that for each sample the impurity concentration $n_i$ is the same both at $B=0$ and at half-filling, this allows us to fit for the conductivity of the composite Fermi liquid with a single Dirac Fermi velocity $v = \sqrt{v_xv_y}=1.8 \times 10^6$ cm/s.  This corresponds to $r_s \sim 10$, which is more than an order of magnitude more strongly interacting than most other condensed matter realizations of Dirac fermions.  Finally, we encourage experimentalists working with graphene, twisted bilayer graphene, surface states of topological insulators or layered organic conductors (all of which support anisotropic bare Dirac fermions) to look for our proposed universal many-body renormalization induced suppression of the anisotropy. \\

\matmethods{The methods are described in the supplementary information.
}

\showmatmethods{} 

\acknow{We acknowledge several discussions with João Rodrigues and Pinaki Sengupta with whom we worked closely on a separate, but related project.  It is a pleasure to thank Bertrand Halperin, Jainendra Jain and Mansour Shayegan for useful suggestions.  We also thank Miguel Dias Costa for assistance with the numerical parallelization, Wei Pan for providing us with the experimental data in Fig. \ref{fig:FigVelocity}, and Xingyu Gu for discussions.  The work was made possible by allocation of computational resources at the CA2DM (Singapore) and the Gauss Centre for Supercomputing (SuperMUC at the  Leibniz Supercomputing Center), and funding by the Singapore Ministry of Education (MOE2017-T2-1-130), Deutsche Forschungsgemeinschaft (SFB 1170 ToCoTronics, project C01). MT acknowledges support of the CA2DM Director's Fellowship (NRF Medium Sized Centre Programme R-723-000-001-281).}

\showacknow{} 

\bibliography{pnas-sample,Ref}

\end{document}